\documentclass[aps,prb,floatfix,superscriptaddress,showpacs,showkeys]{revtex4}
\usepackage{amsmath}
\usepackage{graphicx}
\usepackage{amssymb}

\input{tcilatex}

\begin{document}

\title{Spin-orbit coupling and magnetic spin states in cylindrical quantum
dots}
\author{C. F. Destefani}
\affiliation{Department of Physics and Astronomy, Ohio University, Athens, Ohio 45701-2979}
\affiliation{Departamento de F\'{\i}sica, Universidade Federal de S\~{a}o Carlos,
13565-905, S\~{a}o Carlos, S\~{a}o Paulo, Brazil}
\author{Sergio E. Ulloa}
\affiliation{Department of Physics and Astronomy, Ohio University, Athens, Ohio 45701-2979}
\author{G. E. Marques}
\affiliation{Departamento de F\'{\i}sica, Universidade Federal de S\~{a}o Carlos,
13565-905, S\~{a}o Carlos, S\~{a}o Paulo, Brazil}

\begin{abstract}
We make a detailed analysis of each possible spin-orbit coupling of
zincblende narrow-gap cylindrical quantum dots built in a two-dimensional
electron gas. These couplings are related to both bulk (Dresselhaus) and
structure (Rashba) inversion asymmetries. We study the competition between
electron-electron and spin-orbit interactions on electronic properties of
2-electron quantum dots.
\end{abstract}

\maketitle

The creation and manipulation of spin populations in semiconductors has
received great attention since the Datta-Das proposal of a spin field-effect
transistor,\cite{1} based on Rashba spin-orbit coupling of electrons in a
bidimensional electron gas,\cite{2} and the possibility for quantum
computation devices using quantum dots (\textbf{QDs}).\cite{3} Thus, it is
important that every spin-orbit (\textbf{SO}) effect be clearly understood
for a full control of spin-flip mechanisms in nanostructures.

There are two main \textbf{SO} contributions in zincblende materials. In
addition to the structure inversion asymmetry (\textbf{SIA}) caused by the
2D confinement (the Rashba \textbf{SO}), there is a bulk inversion asymmetry
(\textbf{BIA}) term in those structures (the Dresselhaus \textbf{SO}).\cite%
{5} An yet additional lateral confinement defining a dot introduces another
\textbf{SIA} term with important consequences, as we will see in detail.
Although the relative importance of these two effects depends on the
material and on sample design (via interfacial fields), only recently have
authors begun to consider the behavior of spins under the influence of all
effects.

The goal of this work is to show how important different types of \textbf{SO}
couplings are on the spectra of parabolic \textbf{QD}s built with narrow-gap
zincblende materials. We consider both Rashba and a diagonal \textbf{SIA},
as well as the all Dresselhaus \textbf{BIA} terms in the Hamiltonian, in
order to study features of the spectrum as function of magnetic field, dot
size, and electron-electron interaction.

Consider a heterojunction or quantum well confinement potential $V(z)$ such
that only the lowest $z$-subband is occupied. The Hamiltonian for a
cylindrical \textbf{QD}, in the absence of \textbf{SO} interactions, is
given by $H_{0}=(\hbar ^{2} /2m)\mathbf{k}^{2}+V(\rho )+g\mu _{B}\mathbf{B}%
\cdot \mathbf{\sigma} /2$, where $\mathbf{k}=-i\mathbf{\nabla}+e \mathbf{A}%
/(\hbar c)$, $\mathbf{A}=B\rho (-\sin \theta ,\cos \theta ,0)/2$ describes a
magnetic field $\mathbf{B}=B\mathbf{z}$, $m$ is the effective mass in the
conduction band, $g$ is the bulk $g$-factor, $\mu_{B}$ is Bohr's magneton, $%
V(\rho)=m\omega_{0}^{2}\rho^{2} /2$ is the lateral dot confinement, and $%
\mathbf{\sigma}$ is the Pauli spin vector. The analytical solution of $H_{0}$
yields the Fock-Darwin (\textbf{FD}) spectrum, $E_{nl\sigma_{Z}}=(2n+|l|+1)%
\hbar \Omega +l\hbar \omega_{C}/2+g\mu _{B}B\sigma_{Z}/2$, with effective
(cyclotron) frequency $\Omega =\sqrt{ \omega_{0}^{2}+\omega_{C}^{2}/4}$ ($%
\omega_{C}=eB/(mc)$). The \textbf{FD} states are given in terms of Laguerre
polynomials.\cite{FD} The lateral, magnetic and effective lengths are $l_{0}=%
\sqrt{\hbar /(m\omega_{0})}$, $l_{B}=\sqrt{\hbar /(m\omega_{C})}$ and $%
\lambda =\sqrt{\hbar /(m\Omega)}$, respectively.

The \textbf{SIA} terms\cite{2} for the full confining potential, $V(\mathbf{r%
})=V(\rho)+V(z)$, and coupling parameter $\alpha$ are decomposed as $%
H_{SIA}=H_{R}+H_{SIA}^{D}+H_{K}$. These three forms are: i) $H_{K}=i\alpha
(\hbar \omega_{0}/l_{0}^{2})\lambda
x[\sigma_{+}L_{-}-\sigma_{-}L_{+}]\left\langle k_{z}\right\rangle $ gives
zero contribution when $\left\langle k_{z}\right\rangle =0$ (pure state
parity); ii) $H_{SIA}^{D}=\alpha \sigma_{Z}(\hbar \omega
_{0}/l_{0}^{2})[L_{Z}+\lambda^{2}x^{2}/2l_{B}^{2}]$ is the diagonal
contribution due to the lateral confinement and $x=\rho /\lambda $ been a
dimensionless radial coordinate, $L_{Z}=-i\partial /\partial \theta$ is the $%
z$-orbital angular momentum; iii) $H_{R}=-\alpha (dV/\lambda
dz)[\sigma_{+}L_{-}A_{-}+\sigma_{-}L_{+}A_{+}]$ is the Rashba term for the
perpendicular confinement $dV/dz$, $L_{\pm}=e^{\pm i\theta }$, $\sigma_{\pm
}=(\sigma_{X}\pm i\sigma_{Y})/2$ and $A_{\pm }=[\mp \partial /\partial
x+L_{Z}/x+x\lambda ^{2}/(2l_{B}^{2})]$. In principle these terms can be
tuned since $H_{SIA}^{D}$ depends on the confining frequency $\omega_{0}$
while $H_{R}$ depends on the interfacial field $dV/dz$.

The \textbf{BIA} Hamiltonian \cite{5} for zincblende materials, after
averaging along the $z$-direction, is given by $H_{BIA}=\gamma \left[ \sigma
_{x}k_{x}k_{y}^{2}-\sigma _{y}k_{y}k_{x}^{2}\right] +\gamma \left\langle
k_{z}^{2}\right\rangle \left[\sigma_{y}k_{y}-\sigma _{x}k_{x}\right] +\gamma
\sigma_{z}\left\langle k_{z}\right\rangle \left( k_{x}^{2}-k_{y}^{2}\right)$%
, where $\gamma$ is the Dresselhaus parameter. The first (second) term is
cubic (linear) in the in-plane momentum. The last term will be zero for
systems where $\left\langle k_{z}\right\rangle =0$, while $\left\langle
k_{z}^{2}\right\rangle \simeq (\pi /z_{0})^{2}$, $z_{0}$ being the $z$%
-direction (perpendicular) confinement length. In cylindrical coordinates $%
H_{BIA}$ can be written as $H_{BIA}=H_{D}^{L}+H_{D}^{C}$, where $%
H_{D}^{L}=-i(\gamma /\lambda )\left[ \sigma_{+}L_{+}A_{+}-\sigma
_{-}L_{-}A_{-}\right] \left\langle k_{z}^{2}\right\rangle $ is the linear
term and, after long algebra manipulation,\cite{PRB} the cubic term becomes $%
H_{D}^{C}=i(\gamma /\lambda^{3})\left[ \sigma _{-}L_{+}^{3}H_{1}+\sigma
_{+}L_{-}^{3}H_{2}+\sigma_{-}L_{-}H_{3}+\sigma _{+}L_{+}H_{4}\right] $,
where $H_{i}=A_{i}+\frac{\lambda^{2}}{l_{B}^{2}} B_{i}+\frac{\lambda ^{4}}{%
l_{B}^{4}}C_{i}+\frac{\lambda^{6}}{l_{B}^{6}} D_{i} $, with $i=1,2,3,4$. The
long expressions for the sixteen functions $A_{i}$, $B_{i}$, $C_{i}$, $D_{i}$
are given in Ref. [\onlinecite{PRB}].

Finally the electron-electron interaction $H_{ee}=e^{2}/[\varepsilon |
\mathbf{r}_{1}-\mathbf{r}_{2}|]$, with $\varepsilon $ being the dielectric
constant of the material, is expanded into Bessel functions $J_{k}(\xi )$ as
$H_{ee}=(\hbar \Omega \lambda /a_{B})\sum_{k=-\infty }^{\infty }e^{ik(\theta
_{1}-\theta _{2})}\int_{0}^{\infty }d\xi J_{k}(\xi x_{1})J_{k}(\xi
x_{2})e^{-\xi z_{0} /\lambda}$, where $a_{B}=\varepsilon \hbar ^{2}/(me^{2})$
is the effective Bohr radius. The \textbf{FD} basis states must be properly
antisymmetrized to describe unperturbed spin eigenstates.\cite{PRB}

Summarizing, our total single-particle Hamiltonian is given by $%
H=H_{0}+H_{SIA}^{D}+H_{R}+H_{D}^{L}+H_{D}^{C}$. For the two-particle case,
we study the states and spectrum of $H+H_{ee}$. Parameters for InSb are in
Ref. [\onlinecite{parameters}].

We present results by analyzing the role of each \textbf{SO} term in the
Hamiltonian. We take into account all states in the \textbf{FD} basis having
$n\leq 4$ and $\left\vert l\right\vert \leq 9$ in our numerical
diagonalization, which is equivalent to the first ten energy shells at zero
field and embodies a total of $110$ basis states. The sequence of \textbf{FD}
states of $H_{0}$ starts at zero $B$-field with $\{n,l,\sigma _{Z}\}\equiv
\{0,0,\pm 1\}$, followed by the degenerate set $\{0,-1,\pm 1\}$ and $%
\{0,1,\pm 1\}$. The next energy shell is composed by $\{0,-2,\pm 1\}$, $%
\{1,0,\pm 1\}$, and $\{0,2,\pm 1\}$.\cite{FD} Spin and orbital degeneracies
are broken by $B$ and the states with negative $l$ and positive $\sigma_{Z}$
acquire lower energies because of the negative $g$-factor of InSb. The
lowest \textbf{FD} crossing occurs between states $\{0,0,-1\}$ and $%
\{0,-1,+1\}$, at a critical field $B_{C}^{0}=\widetilde{m}\hbar
\omega_{0}/[\mu_{B} \sqrt{\widetilde{m}|g|(\widetilde{m}|g|+2)}]$. The
moderate value of $B_{C}^{0}$ in InSb ($\simeq 2.6 $ T for $\hbar \omega
_{0}=15$ meV) is a direct consequence of its large $|g| $ factor.\cite%
{parameters} For GaAs ($|g|=0.44$, $\widetilde{m}=m/m_{0}=0.067$), this
level crossing would appear only at $B_{C}^{0}(GaAs)\simeq 9.5$ T for a much
smaller confinement, $\hbar \omega_{0}=2$ meV, corresponding to a regime
where Landau levels are well defined.

Any figure showing spectrum has the structure: Panel $A$ shows \textbf{QD}
spectrum for the full \textbf{FD} basis ($110$ states); Panel $B$ shows a
zoom on the three lowest shells, plus inset with another zoom on the $4$
levels of the second shell; Panel $C$ and $D$ show, respectively, the $B$%
-evolution of spin $\sigma_{Z}$ and orbital $l$ angular momenta, for the
full \textbf{FD} basis, while their insets take into account only the lowest
$7$ \textbf{QD} levels.

Figure 1 shows the simultaneous addition of both \textbf{SIA} terms, $%
H_{SIA}^{D}$ and $H_{R}$, to $H_{0}$. The diagonal term $H_{SIA}^{D}$ causes
small splittings on the zero-field spectrum and readjust the sequence of
states according to total angular momentum $j=l+\sigma _{Z}/2$. For example,
the highest (lowest) zero-field energy level in the second shell has $j=3/2$
($j=1/2$). Since $H_{SIA}^{D}$ does not induce shift on the accidental
degeneracy points of the \textbf{FD} spectrum at finite fields, the first
level crossing occurs at a critical value $B_{C}\simeq B_{C}^{0}\simeq 2.6$
T. Also, this term does not induce any level mixture on the \textbf{FD}
states. The Rashba term $H_{R}$ introduces a strong state mixture for any
magnitude of $\alpha $. This is evident for any pair of \textbf{FD} levels
satisfying $\Delta l=-\Delta \sigma _{Z}/2=\pm 1$ that show a crossing at
given accidental degeneracy of the \textbf{FD} spectrum. The induced mixture
converts this crossing at $B_{C}^{0}$ to an anticrossing (\textbf{AC}), at a
shifted critical field $B_{C}\simeq 2.5$ T $\lesssim B_{C}^{0}$, with an
energy minigap. Higher energy levels, also satisfying this selection rule,
present \textbf{AC}s around the same value of $B_{C}$, and gives origin to
the observed collapse in both $\sigma _{Z}$ and $l$ quantum numbers at $%
B\simeq 2.5$ T, shown in Panels $C$ and $D$. The range of critical fields
(between $2.1$ and $2.6$ T), and the size of the minigaps opened at those
\textbf{AC} regions, are proportional to the magnitude of $\alpha $. $H_{R}$
also induces small splittings in the zero-field spectrum and slightly shifts
the accidental degeneracy points at finite fields. After adding both \textbf{%
SIA} terms simultaneously (full spectrum in Panel $A$), one can see in the
inset of Panel $B$ that the ordering of states is the one determined by $%
H_{SIA}^{D}$. However, the energies of states $j=3/2$ at $30$ meV and $j=1/2$
at a slightly smaller value, as well as the value of field ($B\gtrsim 0.1$
T) where the normal state ordering (the one existing in the absence of SO
terms) for $g<0$ material is restored, are determined by $H_{R}$. For
increasing energy, the ordering for this second shell is $\{0,-1,+1\}$, $%
\{0,-1,-1\}$, $\{0,1,+1\}$, $\{0,1,-1\}$. The width of critical fields
becomes wider, between $2.2$ and $3.6$ T, as seen in Panel $C$. Furthermore,
Panel $D$ shows that orbitals having $l<0$ ($l>0$) present \textbf{AC}s at
fields smaller (larger) than the field $B_{C}\simeq 2.55$ T where occurred
the first \textbf{AC} (see insets of Panels $C$ and $D$). For a future
comparison notice, in Panel $B$, that the first \textbf{AC} near $50$ meV
involving $n=1$ states $\{0,1,-1\}$ and $\{1,0,+1\}$ occurs at that same $%
B_{C}$ value. In general, \textbf{AC}s between states with any $n$ value
occur inside the same unique range of critical fields, as shown in Panels $C$
and $D$. Finally, observe that both \textbf{SIA} terms can be reduced by
decreasing $\omega _{0}$ ($H_{SIA}^{D}$) or $dV/dz$ ($H_{R}$), and that all
even and negative $l$ states ($l=-2,-4,-6,-8$) show anticrossings. \FRAME{%
ftbphFU}{8.6701cm}{6.0231cm}{0pt}{\Qcb{Spectrum when $H_{SIA}^{D}$ and $%
H_{R} $ are added to $H_{0}$ ($A$ and $B$). Critical field range for \textbf{%
AC}s is seen on $C$ and $D$. The lowest one ($B_{C}\simeq 2.55$ T, insets)
occurs near $B_{C}^{0}$. \textbf{AC}s involving $l<0$ ($l>0$) orbitals are
shifted to lower (higher) fields ($D$).}}{\Qlb{fig1}}{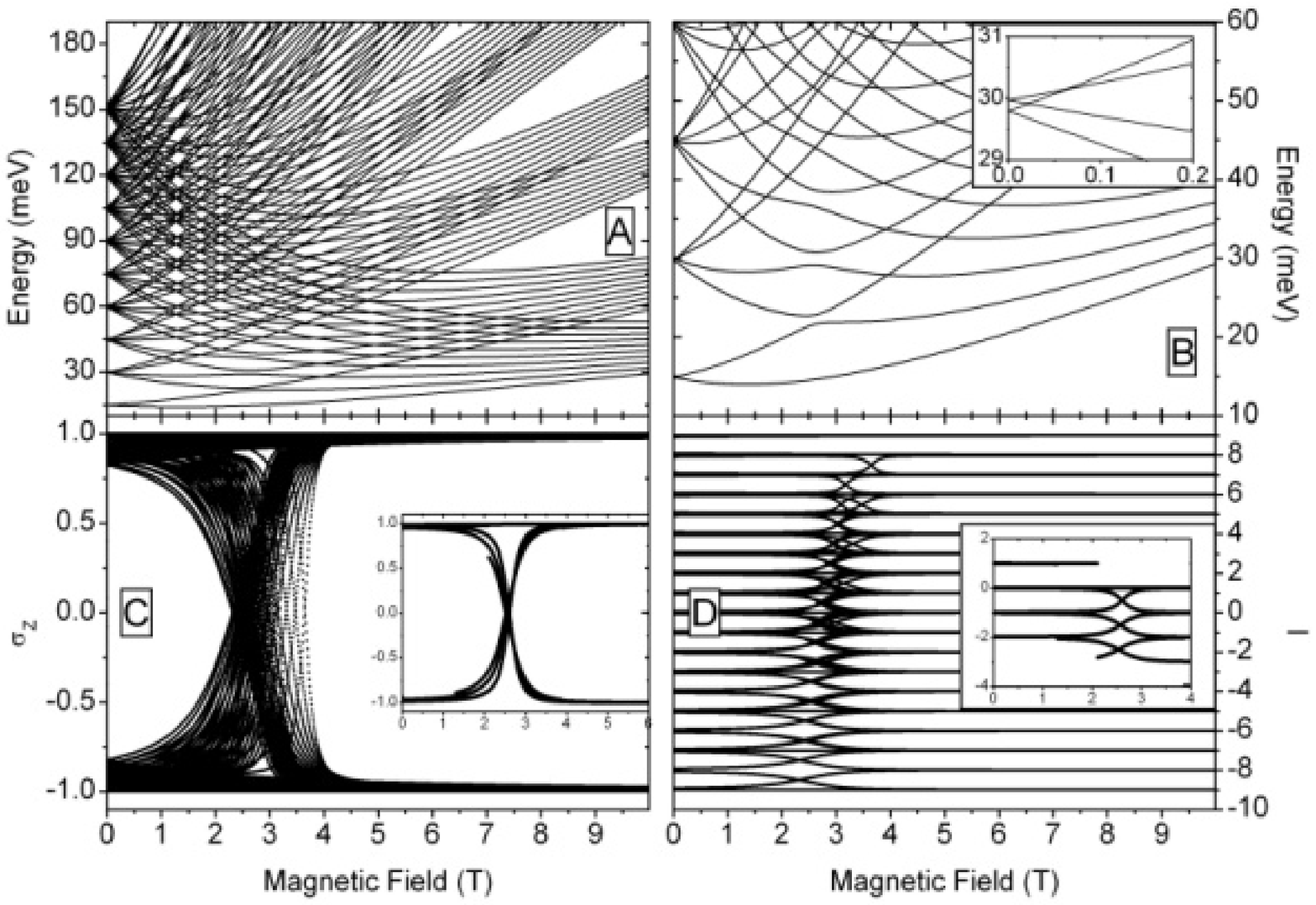}{\special%
{language "Scientific Word";type "GRAPHIC";maintain-aspect-ratio
TRUE;display "ICON";valid_file "F";width 8.6701cm;height
6.0231cm;depth 0pt;original-width 35.4167in;original-height
24.6039in;cropleft "0";croptop "1";cropright "1";cropbottom
"0";filename 'fig1.eps';file-properties "XNPEU";}}

Figure 2 shows the simultaneous addition of both \textbf{BIA} terms, $%
H_{D}^{C}$ and $H_{D}^{L}$, to $H_{0}$. The cubic term $H_{D}^{C}$, under
the present \textbf{QD} parameters, has small influence over the $H_{0}$
spectrum. Small state mixtures is induced at $B_{0}\simeq 1$ and $\simeq 5$
T, both involving \textbf{AC}s satisfying $\Delta l=\mp 3$ and $\Delta
\sigma _{Z}/2=\pm 1$. The first one, at $1$ T ($5$ T), occurs between states
$\{0,1,-1\}$ and $\{0,-2,+1\}$ ($\{0,0,-1\}$ and $\{0,-3,+1\}$). This term
also induces zero-field splittings on the \textbf{FD} spectrum and also a
shift due to matrix elements for $\Delta l=\pm 1=\Delta \sigma _{Z}/2$.
However, such splittings and opened minigaps at the \textbf{AC}s regions are
very small. Therefore, the simultaneous addition of both \textbf{BIA} terms
(full spectrum in Panel $A$), where the linear term $H_{D}^{L}$ is the most
important, drastically changes the general features of the \textbf{FD}
spectrum, inducing strong zero-field splittings and shifting its accidental
degeneracies to higher fields. Yet, the respective matrix elements ($\Delta
l=\pm 1=\Delta \sigma _{Z}/2)$ does not introduce \textbf{AC}s\textbf{\ }on
the lowest energy levels. As seen in Panels $C$ and $D$, the mixing induced
by the linear term is so strong that the \textbf{QD} states are not anymore
pure states even at zero field. Notice in Panel $C$ that at $B=0$ values of $%
\left\vert \sigma _{Z}\right\vert <0.5$ are found for high energy states,
while in its inset one finds $\sigma _{Z}\simeq 0.7$ for the ground state.
As an example of level crossings displaced to higher fields, observe in
Panel $B$ that the first one has moved to $B_{C}\simeq 3.3$ T, there is only
one crossing present in the second shell at about $0.45$ T (see inset), and
the second one occurs a higher field around $3.5$ T. Thus, contrary to the
observed for \textbf{SIA} case, the normal ordering of state is no longer
restored. We will come back to this fact later. As a final note, in the same
inset and at zero field, the highest (lowest) energy state has $j=3/2$ with
eigenvalue equal to energy of $30$ meV of the pure $H_{0}$ ($j=1/2$ at
smaller energy near $27$ meV). The influence of $H_{D}^{L}$ on the spectrum
changes with $z_{0}$. \FRAME{ftbphFU}{8.6701cm}{6.0231cm}{0pt}{\Qcb{Spectrum
when $H_{D}^{C}$\ and $H_{D}^{L}$ are added to $H_{0}$. The linear
contribution dominates the cubic one. Strong mixing at low fields are due to
$H_{D}^{L}$, while the \textbf{AC} around $6$ T is due to $H_{D}^{C}$.}}{%
\Qlb{fig2}}{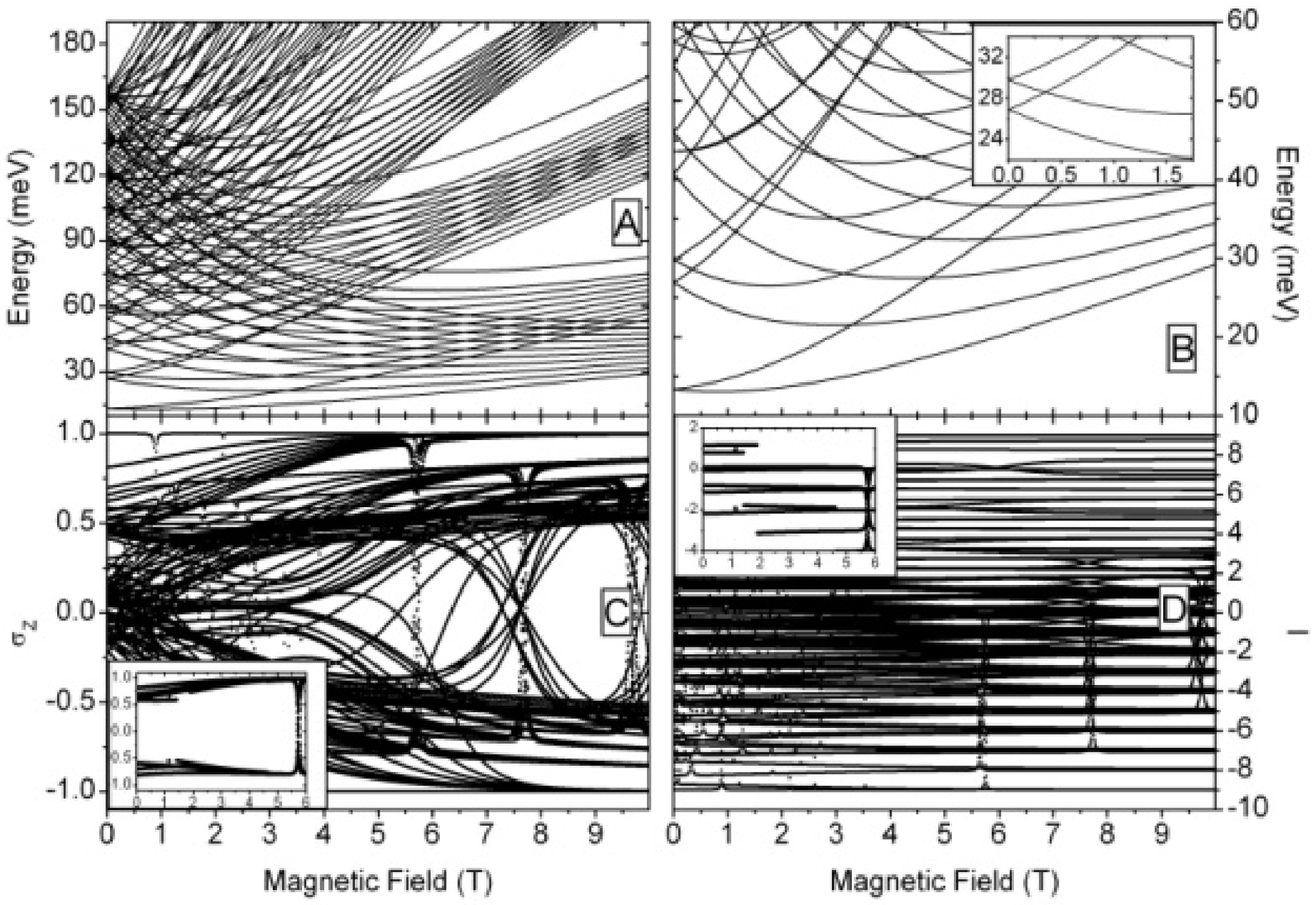}{\special{language "Scientific Word";type
"GRAPHIC";maintain-aspect-ratio TRUE;display "ICON";valid_file
"F";width 8.6701cm;height 6.0231cm;depth 0pt;original-width
35.4167in;original-height 24.6039in;cropleft "0";croptop
"1";cropright "1";cropbottom "0";filename
'fig2.eps';file-properties "XNPEU";}}

Figure 3 shows the one-particle \textbf{QD} spectrum for full $H$ or when
all \textbf{SO} terms are simultaneously taken into account. From the
previous discussions, one may identify which \textbf{SO} mechanism is
dominant in each of the main signatures present on the spectrum. An enormous
state mixture, even at small magnetic fields (Panels $C$ and $D$ and their
insets), as also splittings, position and ordering of states (Panels $A$, $B$
and its inset) are dominated by $H_{D}^{L}$, although with contributions
from both \textbf{SIA}\ terms. The small influence of $H_{D}^{C}$ remains
around $6$ T. The lowest \textbf{AC}s are induced by the Rashba term $H_{R}$%
, although shifted to higher critical fields by the linear \textbf{BIA}
term, $H_{D}^{L}$. Observe that the first \textbf{AC} has moved from $2.55$
T (Fig. 1) to $3.3$ T (Panel $B$ and insets of Panels $C$ and $D$ ), and the
ranges of critical fields becomes wider. However, the new feature of the
full $H$ spectrum is the clear presence of more than one unique range of
critical fields where \textbf{AC}s occur (compare with Fig. 1). Details on
Panels $A$ and $C$: i) The first family of \textbf{AC}s, near $3.3$ T
(related to states between $20$ and $70$ meV) involves only $n=0$ levels.
The first \textbf{AC}s being between $\{0,0,-1\}$ and $\{0,-1,+1\}$, $%
\{0,-1,-1\}$ and $\{0,-2,+1\}$, $\{0,-2,-1\}$ and $\{0,-3,+1\}$, ... ; ii) A
second family of \textbf{AC}s around $5$ T ( related to states between $70$
and $120$ meV) involves only $n=1$ levels. The first \textbf{AC}s being
between $\{0,1,-1\}$ and $\{1,0,+1\}$, $\{1,0,-1\}$ and $\{1,-1,+1\}$, $%
\{1,-1,-1\}$ and $\{1,-2,+1\}$, ... ; iii) A third family of \textbf{AC}s
around $8$ T (related to states between $130$ and $180$ meV) involves only $%
n=2$ levels. The first \textbf{AC}s being between $\{0,2,-1\}$ and $%
\{1,1,+1\}$, $\{1,1,-1\}$ and $\{2,0,+1\}$, $\{2,0,-1\}$ and $\{2,-1,+1\}$,
... . Although lowest \textbf{AC}s in the \textbf{QD} spectrum are caused by
the selection rules of $H_{R}$, the presence of $H_{D}^{L}$ and $H_{D}^{C}$,
in the full $H$, displaces and regroup all states with same $n$ value that
contribute to the minigap near a fixed critical field value.\FRAME{ftbphFU}{%
8.6701cm}{6.0231cm}{0pt}{\Qcb{Spectrum of full $H$, where $H_{R}$ induces
minigap regions that are shifted to higher fields by $H_{D}^{L}$. $H_{D}^{C}$
has small influence but induces state mixtures. The zero-field splittings
produced by $H_{SIA}^{D}$ are dominated by those from $H_{D}^{L}$.}}{\Qlb{%
fig3}}{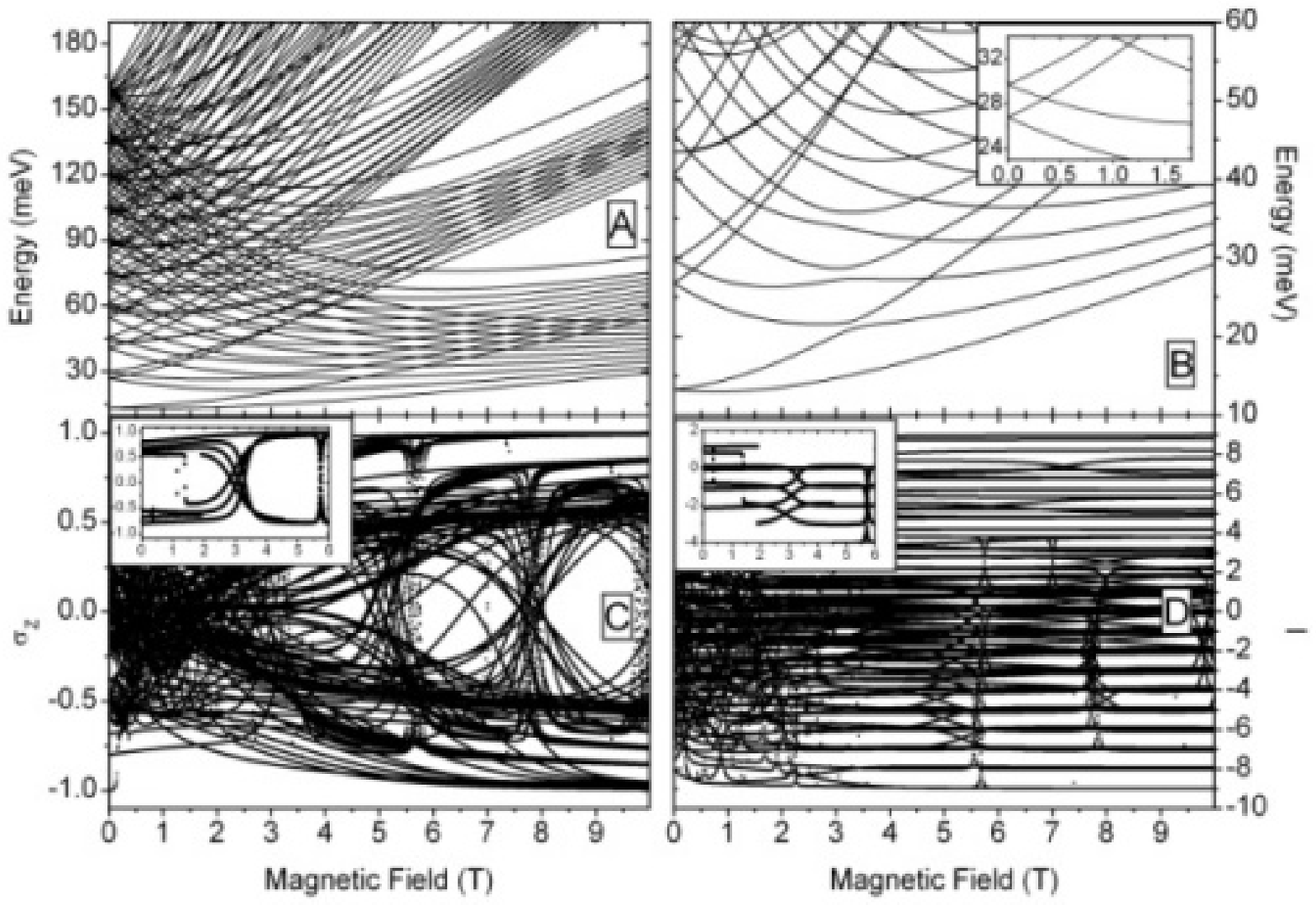}{\special{language "Scientific Word";type
"GRAPHIC";maintain-aspect-ratio TRUE;display "ICON";valid_file
"F";width 8.6701cm;height 6.0231cm;depth 0pt;original-width
35.4167in;original-height 24.6039in;cropleft "0";croptop
"1";cropright "1";cropbottom "0";filename
'fig3.eps';file-properties "XNPEU";}}

In Fig. 4 we simulate the cancellation of zero-field splittings even in the
presence of all \textbf{SO} terms, what is reasonably obtained by taking an
interfacial field $dV/dz$ four times stronger than that one considered
before (see Ref.[\onlinecite{parameters}], other parameters remained
unchanged). This is equivalent to increase the influence of the Rashba term $%
H_{R}$. Notice, in Panels $A$ and $B$, that not only the zero-field
splittings are nearly vanished, but also the Zeeman splittings are
practically suppressed at low fields ($B_{0}\lesssim 1.5$ T). At zero
magnetic field, an energy shell structure identical to the pure $H_{0}$ and
with the same level separation of $15$ meV is formed at displaced energies.
In the inset of Panel $B$ one sees that the energy of $j=3/2$ level is
pushed near $j=1/2$ level, going from $30$ (in Fig.3 B) to $26.5$ meV (in
Fig.4 B). While the zero-field splittings nearly vanish the energy minigaps
increases, as seen in Panel $B$. The rearrangement of electronic levels is
so remarkable that \textbf{AC}s related to the cubic \textbf{BIA} term at $%
1.2$ T become visible (Panel $B$ and insets of Panels $C$ and $D$). The
minigaps at $33$ ($44$) meV involves states $\{0,1,-1\}$ and $\{0,-2,+1\}$ ($%
\{1,0,-1\}$ and $\{0,-3,+1\}$). Even though the electronic levels are less
disperse here than in Fig. 3, the SO-induced state mixture is much more
intense, as can be seen in Panels $C$ and $D$. Between $0$ and $4$ T, most
of the \textbf{QD} levels have $\left\vert \sigma _{Z}\right\vert <0.5$ and
only the ground state has $\sigma _{Z}\simeq 0.7$. As mentioned before, the
insets of Panels $C$ and $D$ show that a strong Rashba term enlarges the
spin-flip region near $B_{C}$.

\FRAME{ftbphFU}{8.6701cm}{6.0231cm}{0pt}{\Qcb{Full $H$ spectrum with four
times stronger $dV/dz$. Notice the cancellation of zero-field and Zeeman
splittings at low fields ($A$, $B$ and inset). New \textbf{AC}s due to $%
H_{D}^{C}$ selection rules occur near $1.2$ T ($B$ and insets in $C$ and $D$%
). The lowest \textbf{AC} is shifted back to $B_{C}=2.7$ T. Notice the
enormous state mixture in $C$ and, at zero field, most states are displaying
$\left\vert \protect\sigma _{Z}\right\vert <0.5$.}}{\Qlb{fig4}}{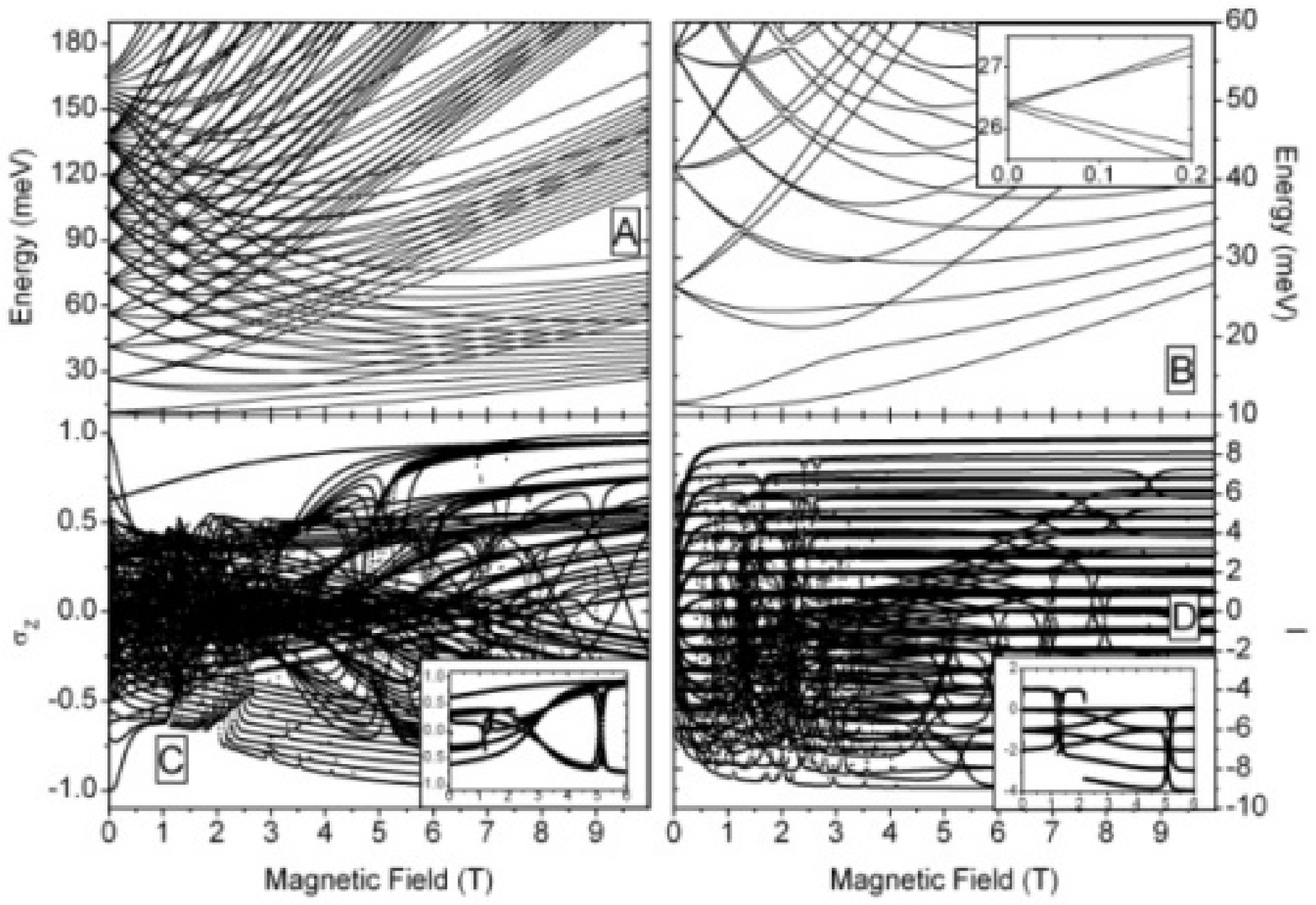}{%
\special{language "Scientific Word";type
"GRAPHIC";maintain-aspect-ratio TRUE;display "ICON";valid_file
"F";width 8.6701cm;height 6.0231cm;depth 0pt;original-width
35.4167in;original-height 24.6039in;cropleft "0";croptop
"1";cropright "1";cropbottom "0";filename
'fig4.eps';file-properties "XNPEU";}}

One can further appreciate the intricate balance between \textbf{SO} terms
by analyzing how some quantities are affected by changes on the lateral and
vertical sizes, $l_{0}$ and $z_{0}$, or on Rashba field, $dV/dz$, as shown
in Fig.~5. Curves with squares, circles and triangles refer to a \textbf{QD}%
, respectively, with parameters of Ref. [\onlinecite{parameters}], with $%
z_{0}$ doubled (smaller linear \textbf{BIA} contribution) and with four
times stronger $dV/dz$, while the dotted curve, in middle Panel, indicates
the $B_{C}^{0}$ field where the first \textbf{FD} level crossing occurs. The
zero-field splitting (left Panel) for states $j=3/2$ and $j=1/2$ of the
second shell is dominated by the linear \textbf{BIA} contribution for any
value of $l_{0}$. An increase on $z_{0}$ strongly reduces the splittings
because the Dresselhaus contribution becomes weaker. The reduction is even
more drastic by increasing $dV/dz$, which makes $H_{R}$ larger and, thus,
may cancel or suppress zero-field splittings produced by $H_{SIA}^{D}$.%
\FRAME{ftbphFU}{8.6679cm}{6.0231cm}{0pt}{\Qcb{Zero-field energy splittings
for the states in the second energy shell (left Panel), critical magnetic
fields where the first level \textbf{AC} (middle Panel) occurs, and energy
minigaps opened at that AC (right Panel) as function of the \textbf{QD}
lateral radius $l_{0}$. Meaning of square, circle and triangle curves are
explained in text. Arrows at $l_{0}=190$ \AA\ show the \textbf{QD} radius
where the spectra from Figs. 1 to 4 were calculated.}}{\Qlb{fig5}}{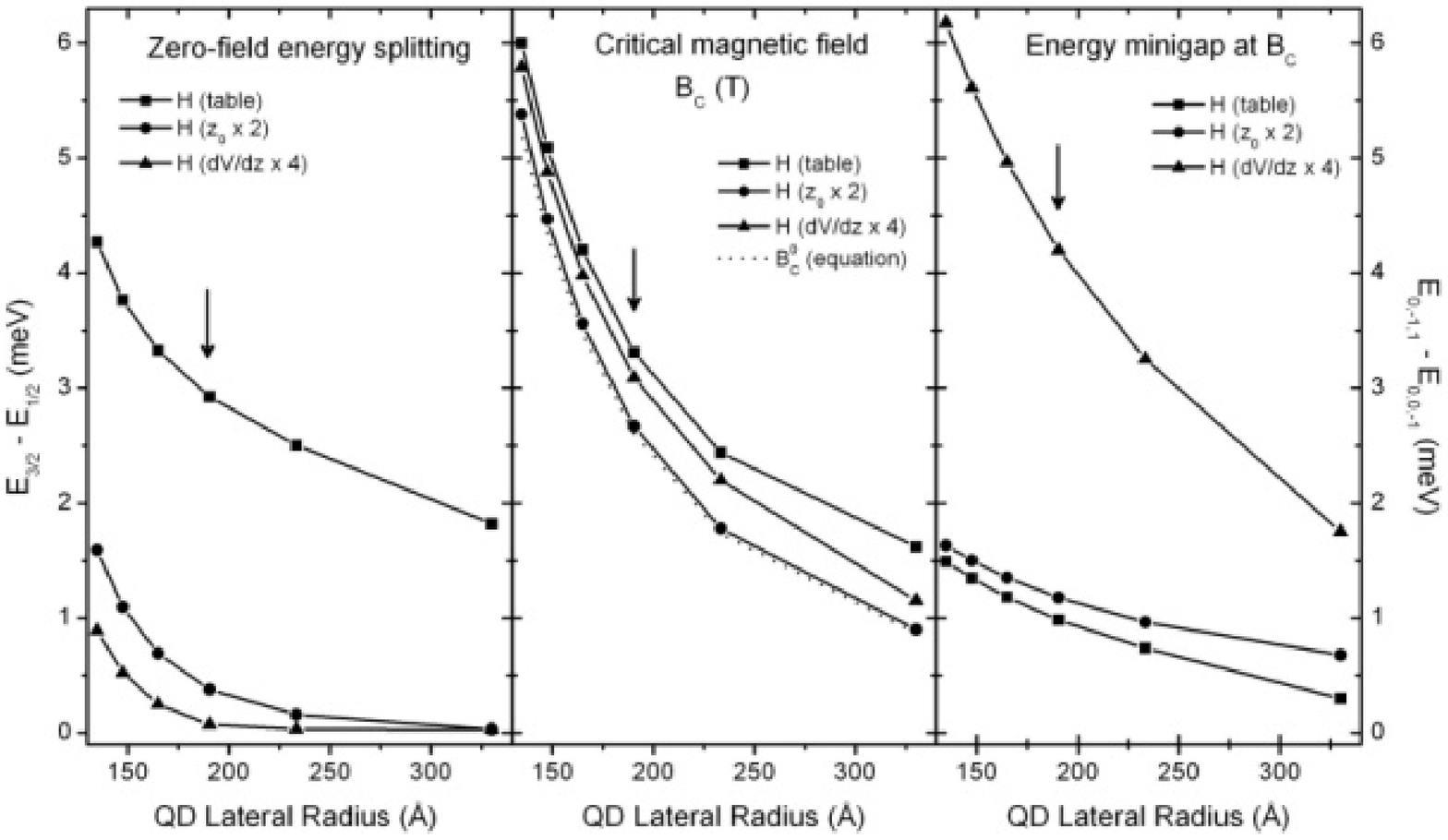}{%
\special{language "Scientific Word";type
"GRAPHIC";maintain-aspect-ratio TRUE;display "ICON";valid_file
"F";width 8.6679cm;height 6.0231cm;depth 0pt;original-width
35.3959in;original-height 24.5935in;cropleft "0";croptop
"1";cropright "1";cropbottom "0";filename
'fig5.eps';file-properties "XNPEU";}}

The critical fields $B_{C}$, where \textbf{AC}s determined by $H_{R}$ occur
(middle Panel, for the lowest minigap between levels $\{0,0,-1\}$ and $%
\{0,-1,+1\}$), decrease with increasing \textbf{QD} size, once $%
B_{C}^{0}\simeq \omega _{0}\simeq 1/\sqrt{l_{0}}$. Its value is close to $%
B_{C}^{0}$ when \textbf{BIA} terms are not present and the inclusion of $%
H_{D}^{L}$ shifts $B_{C}$ to higher value. Increasing $z_{0}$ or $dV/dz$
decreases $B_{C}$, once they will decrease the effects due to $H_{BIA}$. The
value $B_{C}\simeq 2.1$ T, for $l_{0}=270$ \AA\ ($\hbar \omega _{0}=7.5$
meV) is displaced to $1.8$ ($1.5$ T) if $dV/dz$ ($z_{0}$) is four times
larger (doubled). This last value can be compared to reported $B_{C}\simeq
1.7$ T in Ref. [\onlinecite{9}], where \textbf{BIA} terms are absent. The
small difference $\Delta B=0.2$ T can be attributed to the inclusion of
non-parabolicity effects. Anticrossings at such low fields may be
interesting for applications due to easier access.

Finally, the minigap opened at $B_{C}$ (right Panel) has their main origin
in the $H_{R}$ term, while the inclusion of $H_{BIA}$ causes a substantial
reduction. If the value of $z_{0}$ is doubled the splitting is enhanced
slightly. A yet larger $z_{0}$ produces no significant changes. However, the
splitting can be drastically enhanced by increasing the Rashba field as, for
example, changing from $1$ to $4.2$ meV at $l_{0}=190$ \AA , when
interfacial field is increased four times. Measurement of those three
quantities should yield important information on the relative strength of
\textbf{SO} parameters $\alpha $ and $\gamma $.

After having studied the one-particle \textbf{QD }problem we show, in Fig.
6, the two-electron\textbf{\ QD} spectrum under magnetic field (parameters
in [\onlinecite{parameters}]). On the construction of Slater determinant for
two-particle states we use the $20$ lowest one-particle orbitals ($%
\left\vert l\right\vert \leq 3$ and $n\leq 1$), which amounts to $190$
possible two-particle states that can be labeled, in the absence of \textbf{%
SO} interactions, by the projections of orbital ($M_{L}$) and spin ($M_{S}$)
total angular momenta. If no \textbf{SO} term is present in $H$ (Panel $A$),
we verified that the singlet ground state is located at $35$ meV, while the
first excited shell at zero-field is splitted by the exchange interaction,
being composed by a triplet (at $47.5$ meV) and a singlet (at $50$ meV)
states. At very small magnetic field ($\simeq 0.1$ T), the normal sequence
of QD states is restored. For increasing energy and using the notation $%
\left\{ M_{L},M_{S}\right\} $, the ordering is: $\left\{ 0,0\right\} $ for
the ground state, $\left\{ -1,1\right\} $, $\left\{ -1,0\right\} ^{T}$, $%
\left\{ -1,-1\right\} $, $\left\{ 1,1\right\} $, $\left\{ 1,0\right\} ^{T}$,
$\left\{ 1,-1\right\} $ for the first excited triplet ($T$), and $\left\{
-1,0\right\} ^{S}$, $\left\{ 1,0\right\} ^{S}$ for the first excited singlet
($S$). The crossing between ground singlet and first excited triplet states
occurs at $B_{C}^{0(2e)}=2.1$ T.\FRAME{ftbphFU}{8.6701cm}{6.0231cm}{0pt}{%
\Qcb{Two-particle \textbf{QD} spectrum without ($A$) and with ($B$) all
\textbf{SO} terms. It is seen that the \textbf{SO} energy acts against the
direct and in favor of the exchange Coulomb energies. The first excited
triplet, at zero field, is splitted according to the possible $M_{J}$ values
as explained in text. $C$ ($M_{S}$) and $D$ ($M_{L}$) show the lowest ACs as
induced by $H_{R}$ and shifted by $H_{D}^{L}$.}}{\Qlb{fig6}}{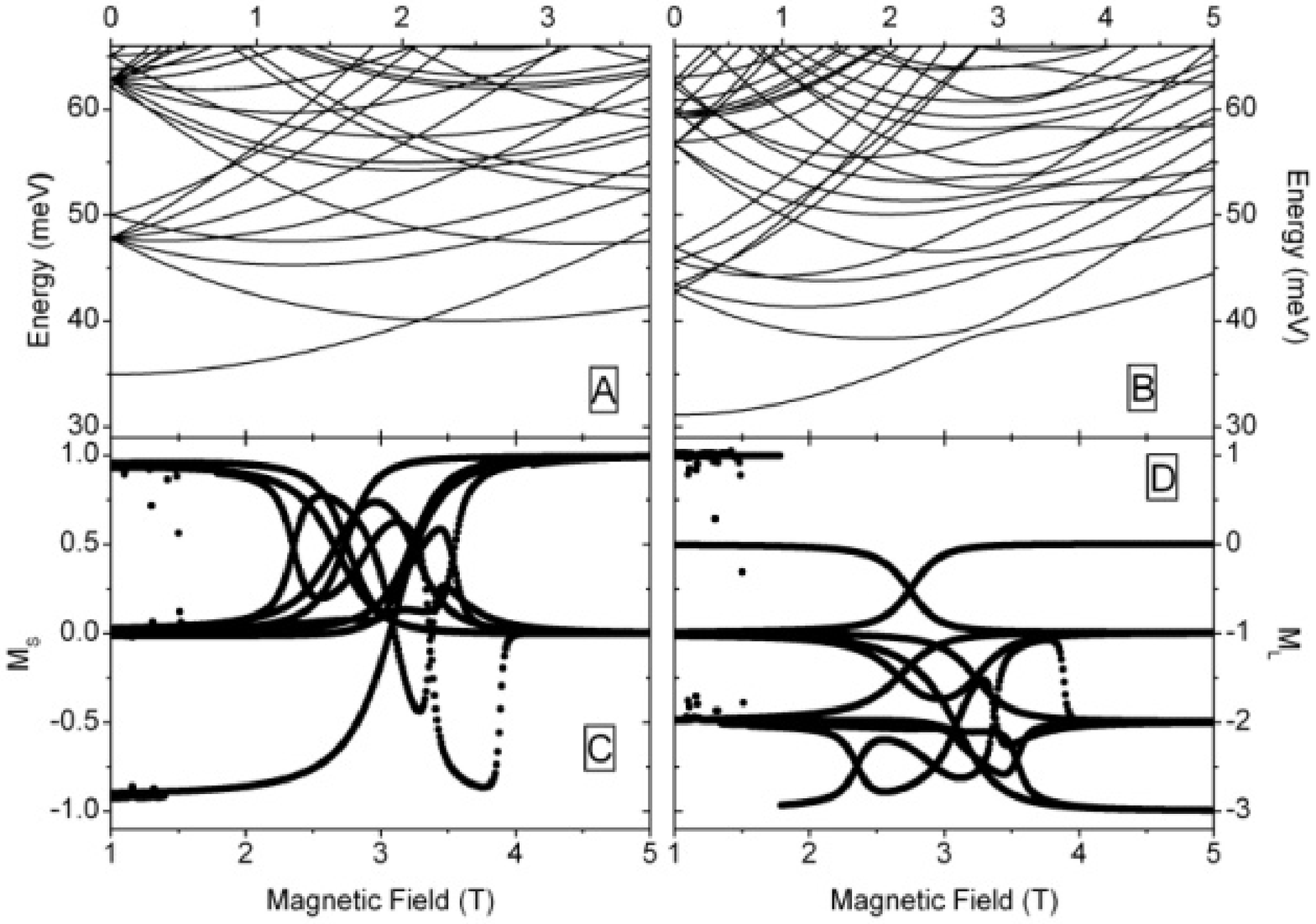}{%
\special{language "Scientific Word";type
"GRAPHIC";maintain-aspect-ratio TRUE;display "ICON";valid_file
"F";width 8.6701cm;height 6.0231cm;depth 0pt;original-width
35.4167in;original-height 24.6039in;cropleft "0";croptop
"1";cropright "1";cropbottom "0";filename
'fig6.eps';file-properties "XNPEU";}}

Panel $B$ shows the \textbf{QD} spectrum for full two-particle Hamiltonian, $%
H_{ee}+H$. We may identify some similar features to the single-particle
case. For example, the linear Dresselhaus term almost destroys the energy
shell structure at zero field by shifting level crossings and inducing new
zero-field splittings, while the Rashba term introduces energy minigaps in
the spectrum. Panel $B$ shows details on the competition between Coulomb and
spin-orbit interactions in narrow-gap cylindrical dots. Observe that the
\textbf{SO} interaction, at zero field, acts against the direct Coulomb
energy and, in a sense, favoring the exchange term. For example: i) The
ground state is shifted back from $35$ to $31$ meV, which is close to the
non-interacting value $30$ meV; ii) The first excited shell states have
energies from $43$ to $47$ meV, values even smaller than the non-interacting
energy $45$ meV. Other important feature in the first excited shell is the
observation that the original triplet is broken on its three possible terms
according to the projection of the total angular momentum, $%
M_{J}=M_{L}+M_{S} $. For increasing energy order, these terms are composed,
at zero field, by the states $\left\{ -1,1\right\} $ and $\left\{
1,-1\right\} $ ($M_{J}=0$), $\left\{ -1,0\right\} ^{T}$ and $\left\{
1,0\right\} ^{T}$ ($\left\vert M_{J}\right\vert =1$), $\left\{ -1,-1\right\}
$ and $\left\{ 1,1\right\} $ ($\left\vert M_{J}\right\vert =2$), in
increasing energy, while the ground ($\left\{ 0,0\right\} $, $M_{J}=0$) and
first excited ($\left\{ -1,0\right\} ^{S}$ and $\left\{ 1,0\right\} ^{S}$, $%
\left\vert M_{J}\right\vert =1$) singlets remain the same.

Panels $C$ and $D$ show the \textbf{SO}-induced mixing of those lowest
states. The first \textbf{AC} at $B_{C}^{(2e)}=2.7$ T involves states $%
\left\{ 0,0\right\} $ and $\left\{ -1,1\right\} $ ($H_{R}$ selection rule
yields $\Delta M_{L}=\pm 1=-\Delta M_{S}$), so that the difference between $%
B_{C}^{(2e)}$ and $B_{C}^{0(2e)}$ ($\sim 0.6$ T) is basically the same
between $B_{C}$ and $B_{C}^{0}$ ($\sim 0.7$ T) for the one-electron problem.
This means that the shifting of $B_{C}$ due to $H_{D}^{L}$ is not altered by
\textbf{QD} occupation, although $B_{C}$ itself is decreased by increased
occupation. Yet, the $H_{D}^{C}$ selection rule becomes $\Delta M_{L}=\mp 3$
and $\Delta M_{S}=\pm 1$, and its first minigap between states $\{-2,1\}$
and $\{1,0\}^{S/T}$ would be visible if a higher $dV/dz$ had been considered
on the solution for the two-electron problem.

A very special difference between the one- and two-particle problems is that
a strong intrinsic (no phonon-assisted) singlet-triplet transition (qubit)
at low magnetic fields involving the ground state becomes possible in the
two-electron case and, in principle, could be explored in implementations of
quantum computing devices. As mentioned, the critical field is decreased by
\textbf{QD} occupation (from $B_{C}=3.3$ to $B_{C}^{(2e)}=2.7$ T), and this
reduction may be increased by decreasing the \textbf{QD} confinement energy.
At these critical fields where the intrinsic state mixture is enhanced, the
\textbf{SO}-induced spin relaxation rate ($\Gamma $) can be estimated from
the minigap energy ($\Delta )$, as $\Gamma =\hbar /\Delta $. For the lowest
\textbf{AC}, $\Delta $ values are taken from the right Panel of Fig. 5, from
where one sees that $\Delta $ is completely changeable by the \textbf{QD}
parameters and, consequently, the intrinsic rate $\Gamma $ can be changed
according to those parameters.

We showed that inclusion of all \textbf{SO} terms is essential in order to
obtain a complete picture of the electronic structure of narrow-gap \textbf{%
QD}s, and discussed the role played by each \textbf{BIA} and \textbf{SIA}
terms on \textbf{QD} spectra and on spin polarization of states. The
combination of strong \textbf{SO} coupling in $H_{R}$ and large $g$-factor
introduces strong intrinsic mixtures and low excitations on the
single-particle spectrum; the position of critical fields where minigaps
occur is affected by $H_{D}^{L}$. We observed that the two-particle spectrum
exhibits strong singlet-triplet coupling involving QD ground state at
moderate fields, which may have significant consequences like possible use
in qubits designs.

Work supported by FAPESP-Brazil, US DOE grant no.\ DE-FG02-91ER45334, and
CMSS Program at OU.

\end{document}